\documentstyle[12pt]{article}
\begin{document}
\begin{titlepage}

\begin{flushright}
UAB-FT-308\\
ILL-(TH)-94-10\\
May 1994
\end{flushright}

\vspace{\fill}

\begin{center}
{\Large Monopole Percolation in pure gauge compact QED}
\end{center}

\vspace{\fill}

\begin{center}
       {\large
        M. Baig and H. Fort\\
        Grup de F\'{\i}sica Te\`orica\\
        and\\
        Institut de F\'\i sica d'Altes Energies\\
        Universitat Aut\`onoma de Barcelona\\
        08193 Bellaterra (Barcelona) SPAIN\\
        \ \\
        and
        \ \\
        J.B. Kogut\\
        Department of Physics\\
        University of Illinois at Urbana-Champaign\\
        1110 West Green Street\\
        Urbana,Il 61801-3080 USA}
\end{center}

\vspace{\fill}

\begin{abstract}
The role of monopoles in quenched compact QED has been studied
by measuring the cluster susceptibility and the order
parameter $n_{max}/n_{tot}$ previously introduced by Hands and Wensley
in the study of the percolation transition observed in non-compact QED.
A correlation between these parameters and the energy (action) at
the phase transition has been observed. We conclude
that the order parameter  $n_{max}/n_{tot}$ is a sensitive probe
for studying the phase transition of pure gauge compact QED.
\end{abstract}

\vspace{\fill}

\end{titlepage}
\newpage

Some time ago Polyakov\cite{p} and Banks, Myerson and Kogut\cite{bmk}
showed that lattice (compact) QED can be written as a
theory of photons and monopoles. This theory exhibits a unique
confining phase for $D=3$ and a two phase structure for $D=4$.
The monopoles play a
central role in the explanation of the phase-structure.
In fact, they produce "disorder" and give raise to
confinement via the {\em dual superconductor}
mechanism \cite{ma}, \cite{tH}.
That is, the gauge vacuum behaves as a `magnetic' superconductor
(a monopole condensate) which
confines the electric field into flux tubes (a dual Meissner effect).

The first numerical simulations studying the effect of monopoles
in pure gauge lattice QED were performed by
DeGrand and Toussaint\cite{dt}.
They searched for monopoles by using Gauss's law, i.e.
measuring the total magnetic flux emanating from a closed surface on the
lattice. In $D=3$ monopoles are pointlike excitations while
in $D=4$ they are closed loops. The density of monopole loops
in four dimensions was measured as a function of the
coupling $\beta$ and a fall-off of the
density across the phase
transition was observed.
This model was also studied by Barber, Shrock and Scharer
\cite{bss}. They simulated the pure gauge compact U(1) theory but
used techniques to suppress monopoles. They concluded that without
monopoles there is no phase transition. This result has been
confirmed by more recent numerical simulations\cite{bmm}.
However, it has proved difficult to establish the order of
the phase transition: initial measurements on small lattices
indicated a second order
transition \cite{ln}
but simulations on larger
lattices which collected greater statistics \cite{jnz}, \cite{acg}
favor a first order transition.

Gupta, Novotny and Cordery\cite{gnc} confirmed that loops of monopole current
are the mechanism driving the phase transition, but conjectured that
the apparent discontinuities in observables might be
due to finite-size effects. In the same spirit,
Grady\cite{g} pointed out the possibility that the "latent heat"
in the phase transition could be a
spurious topological effect due to the imposition of periodic
boundary conditions.
This idea is supported by the recent simulations of
the compact action on  "closed topology" lattices by Lang and
Neuhaus \cite{lne}. In contrast
to studies of compact QED on hypercubic lattices with periodic
boundary conditions, they found no metastability signals at the
phase transition on the lattices with the topology of a sphere.
Numerical simulations with fixed boundary conditions also confirm
the continuity of the transition \cite{bf}.
Additionally,
in a recent paper, Rebbi et al.\cite{rebbi} have
studied the phase structure of pure compact
U(1) lattice action supplemented by a monopole term. They have
found that the strength of the first order transition decreases
with the weight of the added term in such a way that the transition
ultimately gets of second order.

Lattice monopoles can also occur in simpler models. For example,
Kogut, Koci\'c and Hands\cite{kkh} showed that
in pure gauge {\it non-compact} lattice QED monopoles percolate
and satisfy hyperscaling relations characteristic of an
authentic second order phase transition even though the
underlying lattice field theory is trivial.
The non-local character of the monopole observables is an
essential ingredient in having a phase transition in an otherwise
trivial theory. In the
non-compact case dynamical considerations are subsumed by
"geometrical" considerations. As this case illustrates,
percolation is not necessarily connected
with condensation or with other bulk, dynamical
properties of the theory.
Similar phenomena are common in statistical mechanics. For example,
many spin glass models are based on free (Gaussian) dynamics,
but they experience glassy transitions in appropriate non-local
observables which are well studied in laboratory experiments. In
addition, there are models where percolation is not related
to bulk transitions of primary interest.
For example, the 3d Ising model
has no phase transitions at non-zero magnetic field
despite the presence of percolation thresholds.
In addition, a recent study of
scalar lattice QED with noncompact gauge fields\cite{bfkks} showed
that monopole percolation and the bulk Higgs-Coulomb phase transition
occur at separate couplings and are not directly related.

With the aim of obtaining more knowledge about the role of
monopole percolation and condensation phenomena
in lattice quantum electrodynamics,
we have considered the original pure gauge compact  U(1) case, studying
the behaviour of the cluster susceptibility already introduced by
Hands and Wensley\cite{hw}. Our main result is that in compact
pure gauge QED, the percolation threshold occurs at the {\it same}  point
as the deconfining phase transition, i.e. the point of
monopole condensation. Moreover, the $n_{max}/n_{tot}$
order parameter used to
characterize  monopole percolation proves to be an excellent order
parameter to study the location of the
apparent first order phase transition.
The quantity $n_{max}/n_{tot}$ has been advocated recently
in a study of the Villain model of QED
\cite{st}, and, in a different context, in three-dimensional
quantum gravity coupled to gauge fields\cite{rck}.

The lattice action used in our simulation is the original Wilson
(compact) action
\begin{equation}
S_{gauge}=\sum_{n\mu\nu}\cos(\theta_\mu(n)+\theta_\nu(n+\mu)-\theta_\mu(n+\nu)
-\theta_\nu(n))=\sum_{n\mu\nu}\cos \Theta_{\mu\nu}(n).
\end{equation}

Following\cite{dt} we can separate the plaquette angle $\Theta_{\mu\nu}$ into
two pieces: physical fluctuations which lie in the range $-\pi$ to $\pi$ and
Dirac strings which carry $2\pi$  units of flux. Introducing an electric
charge $e$ we define an integer-valued Dirac String by,
\begin{equation}
e\Theta_{\mu\nu}=  e\bar\Theta_{\mu\nu}(\tilde n) + 2\pi S_{\mu\nu},
\end{equation}
where the integer   $S_{\mu\nu}$   determines the strength of the string
threading the plaquette and  $e\Theta_{\mu\nu}$ represents physical
fluctuations. The integer-valued monopole current $m_{\mu}$(n) defined on
links of the dual lattice, is then
\begin{equation}
m_\mu(\tilde n)= {1\over 2} \epsilon_{\mu\nu\kappa\lambda}
\Delta_\nu^+ S_{\kappa\lambda}(n+\hat\mu),
\end{equation}
where $\Delta_\nu^+$ is the forward lattice difference operator, and
$m_\mu$ is the
oriented sum of the $S_{\mu\nu}$ around the faces of an elementary cube. This
definition, which is gauge-invariant, implies the conservation law
$\Delta_\mu^- m_\mu(\tilde n)=0$
which means that monopole world lines form closed loops.

Analogous to ref.\cite{kkh} which considered the non-compact
U(1) model, we have used the constructions and concepts of
percolation in order to clarify the confinement/deconfinement
phase transition in the
compact U(1) model. We also have borrowed the idea of a connected cluster
of monopoles on the dual lattice:  one counts the number of dual sites
joined into clusters by monopole line elements. We also have used the
order parameter $M = n_{max}/n_{tot}$ where $n_{max}$
is the number of sites in the largest cluster and $n_{tot}$ is the total
number of connected sites.  Its associated susceptibility reads,
\begin{equation}
\chi={{<\sum_{n_{min}}^{n_{max}} g_n n^2 - n_{max}^2>}\over {n_{tot}}},
\end{equation}
where $n$ labels the size of a cluster occurring $g_n$ times on the dual
lattice.  In general $n_{min} = 2$, but for monopoles $n_{min} = 4$
because of the conservation law.

We have applied the standard Metropolis algorithm to simulate the compact U(1)
lattice gauge theory with standard periodic boundary conditions.  As
it was mentioned, this system exhibits a first order phase transition.
The lattice sizes used have been from $8^4$ to
$14^4$. We have made measurements of the monopole observables
near the phase transition which occurs very close to $\beta = 1.0$
Two different kinds of measurements have been done, first some simple thermal
cycles on the smaller lattice sizes in order to detect
hysteresis phenomena associated with an apparent first order phase transition.
Secondly, a more detailed analysis on larger lattices in order to confirm
the persistence of the discontinuity in the values of the observables.

In Fig. 1 we show the results of a simultaneous  measurement of the standard
monopole density\cite{dt} $\rho$ and the monopole percolation
parameter $M = n_{max}/n_{tot}$ on a $12^4$ lattice.
Each point is a measurement over 10.000 configurations after discarding
5.000 for thermalization. Note that although both observables show a
fall-off after the transition, the jump in the  M parameter looks
particularly abrupt.

The analysis of the cluster susceptibility $\chi$ is shown in Fig. 2.
for lattice sizes of 6, 8, 10, 12 and 14 lattice units.
The measurements are over
sets of 30.000 iterations at each point after a thermalization of 5.000
iterations.  From this graph one sees a peak in the susceptibility just
beyond the phase transition point. This singular behavior does not in
itself indicate the order of the transition. One must measure the volume
dependence of the peak height to address this point.
In the case of the pure gauge
non compact U(1) theory, a true phase transition was detected,
and critical indices were measured sufficiently accurately that
the transition could be classified as four dimensional
bond percolation\cite{kkh}.
A finite size analysis of our compact case shows that
the peak grows with a different power exponent than the non compact case
but a truly  precise determination of the finite size dependence
was not made. Higher statistics and measurements on a finer grid
of $\beta$ values would be necessary for that.
Our conclusion is that both monopole parameters $n_{max}/n_{tot}$ and
$\chi$ indicate an apparent first order phase transition, just as the
internal energy does.

To confirm this scenario we show in Fig. 3 the evolution
of the internal energy, the cluster susceptibility $\chi$ and the
$n_{max}/n_{tot}$ order parameter in "Monte Carlo time", i.e.
the iteration number, in a set of measurements for a coupling
just beyond the phase transition on a $10^4$ lattice.
The "tunneling" between the two phases is very clear
for all three observables at the same point. This phenomena
is characteristic of a first order phase transition.

As was pointed out recently in \cite{rebbi}, we have
observed that the tunneling between the phases is strongly
suppressed. This fact difficult the study of the phase transition on
larger lattices. Nevertheless we have confirmed the behavior of the
monopole percolation parameters on a $14^4$ lattice.
Since our main objective is
to study the behaviour of monopoles over the phase transition,
we have not applied any iterative reweighting procedure to
locate the phase transition point.
In Fig. 4 we show the evolution of the monopole susceptibility and the
$n_{max}/n_{tot}$ parameter for $\beta=1.01040$ (value very close to the
critical point, but still in the strong coupling phase) starting from an
ordered initial configuration. These results show clearly the change of
the phase, indicating that the parameter
$n_{max}/n_{tot}$ exhibits, greatly amplified, the characteristic
discontinuity of a first order transition.

In conclusion, the analysis performed here shows that the observable
$n_{max}/n_{tot}$, defined by Hands and Wensley \cite{hw} in the
study of the percolation phenomena of non-compact lattice
electrodynamics, proves to be a useful order parameter for the
study of the compact U(1) model phase transition. This
observable has definite advantages when compared with other order
parameters\cite{ipp}, such as the monopole density $\rho$,
since it has a larger "discontinuity" separating the two phases.
Finally, these results show that in the pure gauge compact case
the percolation threshold occurs at the same point as the bulk
phase transition where monopole condensation occurs.

\noindent{\em Acknowledgements}

This work has been partially supported by CICYT (project
number AEN 93-0474). Numerical simulations have been done mainly
on the CRAY-Y/MP of CESCA (Centre de Supercomputaci\'o de Catalunya).
Part of the work has been performed also on the CRAY-Y/MP of CIEMAT (Madrid)
and the CRAY-C90 of the Pitssburg Supercomputer Center (PSC).
J. B. K. is supported in part by the National Science Foundation,
NSF-PHY92-00148.

\newpage

\noindent{\large Figure captions}

\begin{enumerate}

\item  Simultaneous measurements of the monopole density $\rho$ (dashes)
and the monopole percolation $M = n_{max}/n_{tot}$ parameter (solid)
on a $12^4$ lattice.  The statistics is 10.000 iterations at each point
after discarding 5.000 for thermalization.

\item  Measurements of the cluster susceptibility $\chi$ on different lattice
sizes ( squares $6^4$, diamonds $8^4$, crosses $10^4$, circles $12^4$
and stars $14^4$). Lines are only to guide the eye.

\item  Simultaneous measurement of the Internal Energy E, the cluster
susceptibility $\chi$ and the monopole order
parameter $M = n_{max}/n_{tot}$ on
a single run at $\beta = 1.02   $ on a $10^4$ lattice. Each point
is an average over 20 consecutive iterations. The total statistics is
25.000 iterations after discarding 5000 for thermalization.

\item  Results of the measurement of the monopole susceptibility and
 the $M = n_{max}/n_{tot}$ parameter
from a simulation on a $14^4$ lattice at $\beta = 1.01040$.

\end{enumerate}

\end{document}